# Light Enhanced Blue Energy Generation using MoS$_2$ Nanopores

Michael Graf[1*], Martina Lihter[1], Dmitrii Unuchek[2], Aditya Sarathy[3], Jean-Pierre Leburton[3], Andras Kis[2], Aleksandra Radenovic[1*]

[1]Laboratory of Nanoscale Biology, Institute of Bioengineering, School of Engineering, EPFL, 1015 Lausanne, Switzerland

[2]Laboratory of Nanoscale Electronics and Structures, Institute of Electrical Engineering and Institute of Materials Science and Engineering, School of Engineering, EPFL, 1015 Lausanne, Switzerland

[3]Department of Electrical and Computer Engineering, University of Illinois at Urbana-Champaign, USA

[*]All correspondence should be addressed to: michael.graf@epfl.ch and aleksandra.radenovic@epfl.ch

## Summary

**Blue energy relies on the chemical potential difference generated between solutions of high and low ionic strength and would provide a sun-and-wind independent energy source at estuaries around the world. Converting this osmotic energy through reverse-electrodialysis relies on ion-selective membranes. A novel generation of these membranes is based on atomically thin MoS$_2$ membranes to decrease the resistance to current flow to increase power output. By modulating the surface charge by light we are able to raise the ion selectivity of the membrane by a factor of 5 while staying at a neutral pH. Furthermore, we find that the behavior of small nanopores is dominated by surface conductance. We introduce a formalism based on the Dukhin number to quantify these effects in the case of a concentration gradient system. As a consequence, the charges created by light illumination provoke two important changes. Increased surface charge at the pore rim enhances the ion selectivity and therefore larger osmotic voltage (dominating in small pores), while the increased surface charge of the overall membrane enhances the surface conductance and therefore the osmotic current (dominating in larger pores). The combination of these effects might be able to efficiently boost the energy generation with arrays of nanopores with varying pore sizes.**

## Introduction

The term blue energy embodies all the attempts to harvest energy coming from the spontaneous and irreversible mixing of sea-water and river-water. The chemical potential difference between two liquids of different salt concentration holds immense amounts of energy: 2.3MJ of theoretical energy is buried in each cubic meter of water[1]. Extracting this energy can be achieved by reverse-electrodialysis, which relies on the direct electrical conversion of an ion current generated by passing ions through stacks of cation and anion-selective membranes[1]. Ion exchange membranes for reverse electrodialysis are relatively thick and typically made from polymers that contain charged functional groups responsible for rejecting certain types of ions. However, an ideal membrane should be as thin as possible in order to decrease the resistance to current flow and increase the power output. Recent advances in 2D material growth and processing allowed the development free-standing membranes using ultra-thin materials like graphene, or transition metal dichalcogenides (TMDs)[2], such as molybdenum disulfide ($MoS_2$). Controlled drilling in these materials using a transmission electron microscope or an electrochemical reaction[3] allows the creation of single nanopores. Single-nanopores in $MoS_2$ have recently been successfully used to generate nanowatts of power[4]. The relatively strong negative charge of the pore rim and the electrical double layer (EDL) overlap in small pores results in ion repulsion. This combined effects are responsible for an efficient ion-selectivity which manifests in a measured osmotic potential. However, a persistence of high ion-selectivity for relatively big pores (>10nm) has been observed[4,5]. This intriguing effect has not been completely understood so far.

Using single-layer $MoS_2$ we can engineer a reverse-electrodialysis membrane with the highest single pore power to date[4]. However, these highly efficient power conversions have been performed in extremely alkaline conditions (pH 11) to increase the surface charge of the material and improve the ion-selectivity[4,6]. Measurements in such basic environments are not easily transferable to real-world operating conditions, where estuaries and seawater have pH values around 7.4. If these novel membranes pertain to have a winning chance of being applicable in reverse-electrodialysis applications, alternative ways of tuning the surface charge without compromising their thickness need to be developed. For this purpose, another natural resource such as sunlight can be used to generate charges in the material.

The presence of a visible range direct band gap in the single-layer TMDs makes them very attractive for optoelectronics applications[7]. We highlight $MoS_2$ as it shows very strong photogating effects allowing significant modulation of the charge carrier density by illumination with external light[8,9]. This photogating effect is responsible for the relatively slow but ultrahigh sensitivity of $MoS_2$ photodetectors[10].

In this work, we investigate the possibility of tuning the surface charge with light in single-layer $MoS_2$ nanopores. We show that the observed changes in the efficiency of the generated osmotic power are due to photon absorption and the generation of charge carriers within the material. To our knowledge this is the first demonstration of a light-induced efficiency boost for osmotic power harvesting. Furthermore, we optimize the light intensities and the excitation wavelength to maximize the power output, while avoiding any material damage that could be possible due to the process of photo-oxidation[11,12]. The light intensities used fall within ranges that are easily attainable in solar applications. In the end, we go into the analysis of observed high ion-selectivity in big pores. We illustrate that the osmotic potential (and thus ion-selectivity) depends on the interplay between surface conductance and bulk conductance, which in small nanopores results in reduced osmotic potential compared to bigger pores. We present a formalism based on the Dukhin number[13] to estimate the pore-size dependence of the osmotic potential that, for the first time, explains qualitatively the data observed in this paper and in previous studies using high aspect-ratio nanopores[4,5].

## Results

**Experiment.** Figure 1a shows the experimental set-up. We fabricated freestanding $MoS_2$ membranes by transferring CVD grown material on top of an opening in a silicon nitride membrane (typically 50-100nm)[14]. Subsequently, a nanopore is fabricated either by transmission electron microscope (TEM) drilling (Figure 1b) or by the electrochemical reaction (ECR) method[3]. The membranes are then mounted in a flow cell. A concentration gradient of 1mM/100mM KCl is applied between the two compartments to approach real-world values of the concentration of ions in estuaries and ocean[15]. Two Ag/AgCl electrodes connected to a current amplifier measure the ionic current (Figure 1a). We used two diode-lasers at 643nm (red) an 475nm (blue) wavelengths to irradiate the membrane surface during the experiment. The energies of the two excitation wavelengths are above the single-layer $MoS_2$ optical bandgap which is around 1.83 *eV*, i.e. 676nm in wavelength[16]. Specifics on the laser spot size and power measurements can be found in Figure S1 and Table S1. Excitation power densities for the 643nm and 475nm wavelength were calculated to be $P_0*15.05$ W/cm$^2$ and $P_0*27.25$ W/cm$^2$, where $P_0$ was the power set in the control-software.

The lasers were aligned perpendicular to the membrane by adjusting the position of the flow cell to maximize the light transmitted through the membrane (Supplemental Movie M1). We first set out to investigate the pore size stability of the $MoS_2$ membrane. We found that a laser flux of 3 W/cm$^2$ was sufficient to slowly increase the pore size in $MoS_2$ or create additional holes or defects (Figure S2). Therefore, we restricted all of our subsequent laser irradiation to values below 1.5 W/cm$^2$ in order to prevent any misinterpretation of the results due to pore size changes.

**Laser induced surface charge changes.** The osmotic current (the current at zero potential) and the osmotic potential (the potential needed to zero the current) is found by measuring the linear ionic current response in a voltage range of -200mV to 200mV (IV). All reported values were corrected for the contribution of the electrode potential difference (Table S2). Qualitatively, during the laser irradiation, we observe an enhanced osmotic potential (Figure 1c). Increases in osmotic potential are related to larger ion selectivity. This could potentially be caused by stronger surface charges on the rim of the nanopore which enhances the repulsion of Cl$^-$ ions. To cross-check this interpretation, we performed finite element method (FEM) simulations and observe a similar trend when increasing the surface charge of the pore (Figure S3).

Further, we investigate the ionic current response to laser irradiation on-off switching. Light-induced generation of charges in $MoS_2$ is not instantaneous. This can be directly observed in a resulting photo-current in $MoS_2$ photodetectors. When light hits the material a characteristic rise time of the photocurrent is measured. Similarly, when the light is turned off, a decay in the current is observed[10]. The slow part of the photo-response is associated with the photogating effect[8]. Since the ionic current is influenced by the surface charge of the membrane, we expect it to follow a similar trend as the generation of charges. To compare the dynamics of the observed electrical photocurrent to the ion current variations in our system, we record a time trace while the shutter of a 643nm laser (1.5 W/cm$^2$) was alternately opened and closed (Figure 1d). A rapid current jump of ~240pA followed by a slow decay back to the initial current prior to the laser illumination is observed. We can fit the decay to an exponential function and extract the half-life of the current-increase ($t_{raise}$=5-7s) and the current decrease ($t_{fall}$=8-9s). These values are comparable to the reported electrical current measurements ($t_{raise}$=4s and $t_{fall}$=9s)[10].

**Osmotic Power Generation.** Next, we created nanopores of 3nm and 10nm size, respectively, using the ECR method[17]. Dehydration and non-linear current voltage characteristics[18], that complicate accurate pore size estimation, restricted us to a starting pore size of 3nm. The larger 10nm pore was previously determined to be the optimal pore size for power generation[4]. First we estimated the pore size in symmetrical 1M KCl condition by fitting the IV to the conductance model (Figure 2a)[19]. We then performed the experiment in gradient condition by exposing the membrane to laser light. To ensure that all observed differences are due to light irradiation and not to any time-related effect, such as pore size increase or pore clogging, a binary illumination condition (on/off) was used to cycle between a dark state and an illuminated state until we recorded at least 10 IVs for each condition. The obtained values of the osmotic current, potential and power are presented as boxplots in Figure 2b-d. For the small nanopore, the osmotic current is increasing significantly for both wavelengths tested. Similarly, the large nanopore shows a significant increase with the red and blue laser. The median of the osmotic potential of the 3nm pore increase from 73mV in the dark state to 92mV (blue laser) and 98mV (red laser), whereas the 10nm pore did not show any significant difference in osmotic potential. Consistently with previously published work[4,5], we observe an increased osmotic potential for the larger 10nm pore compared to the 3nm one.

Figure 2d shows the power generated by the devices (product of the osmotic current and potential). All power increases were statistically significant. The power increases with laser illumination by 19% (red, d=10nm), 15% (blue, d=10nm), 131% (red, d=3nm) and 71% (blue, d=3nm). These results show a clear relationship between the efficiency of the membrane and the wavelength of the light used to illuminate the membrane. Furthermore, the calculated flux for the red laser (742mW/cm$^2$) is nearly half the value of the blue laser (1'300mW/cm$^2$). It becomes therefore clear that the 643nm laser is substantially more efficient in the generation of optical charges.

**Photoluminescence.** We further investigate this difference in optical response for these two wavelengths and have performed micro-photoluminescence (μ-PL) measurements of $MoS_2$ deposited on the membrane and in the vicinity of the membrane. A bright field image of the area considered for the PL measurements is shown in Figure 3a. The white dashed rectangle depicts the 20x20μm$^2$ area used to generate the PL maps shown in Figure 3c and e with 488-nm and 647-nm excitation wavelength correspondingly. These maps represent spatial distribution of the PL intensity at 665 nm emission wavelength. The observed inhomogeneity of the PL strength over the $MoS_2$ flakes which arises due to the two reasons: First, incident light experiences different interference which depends on the substrate. Thus, there is a strong difference of the light absorption by the $MoS_2$ on top of the 20 nm thick $SiN_x$ membrane compared to the similar crystal deposited on the $SiN_x/SiO_2/Si$ substrate. This effect is clearly visible in the bright field image and leads to the respective difference in μ-PL intensity of $MoS_2$ in and out of the $SiN_x$ membrane (spectra 3 and 4 in Figure 3d, f). Second, the modulation of intensity and position of the μ-PL peak inside the single crystal (spectra 1-2-3 and 5-6 in Figure 3d) is associated with the CVD growth process. Apart from these two side effects, the main results of these measurements are clear when the two maps are compared to each other. The $MoS_2$ PL intensity is 8 times higher when a 647 nm laser is used for excitation. This is consistent with previous reports[20]. Even though $MoS_2$ absorption is lower in the case of the red laser, the energy of incident photons is closer to the excitonic resonance making optical excitation more efficient and optical effects more pronounced, including the photogating effect schematically depicted in Figure 3b. Therefore, efficient optical generation of charges in $MoS_2$ under the 643nm excitation is responsible for the results summarized in Figure 2.

## Discussion

**Heat.** As previously reported by Hwang et al.[21], heat can influence the gradient driven energy conversion by increasing the surface charge density and ionic mobility. The authors see a steep change of the slope of the IV characteristics with changing temperature, whereas in this study we merely see an offset (Figure 1c), suggesting that our mechanism cannot be explained by heat induced conductivity changes. Furthermore, Figure 2c clearly shows an increased effect with the 643nm laser, even though the energy provided is nearly half the energy of the 475nm laser. If the effect would be dominated by heating one would expect a more prominent effect using the more energetic 475nm illumination. In Supplemental Information we provide a detailed and thorough analysis on how heat can influence different parameters in our system such as EDL thickness, surface potential and viscosity. Furthermore, through finite element method (FEM) simulations we estimate a maximum of 4°C of temperature increase in our system (all details are provided in Supplemental Information) which is not enough to cause the observed response in current and voltage. We therefore conclude that heat does not play a relevant role in explaining the experimental data.

**Evidence for increased surface-charge density.** The ionic current measurements performed in this work illustrate an enhancement in osmotic voltage (Figure 1c) during laser irradiation. This indicates that the surface-charge density increases, which is consistent with the results obtained by FEM simulations (Figure S3).

As previously mentioned, the ionic current response to laser switching shows similar behaviour as current generated in $MoS_2$ photodetectors. Half-life of the current-increase ($t_{raise}$=5-7s) and the current decrease ($t_{fall}$=8-9s) are similar as in reported electrical current measurements ($t_{raise}$=4s and $t_{fall}$=9s)[10]. The same dynamic response in the measured ionic current indicates a strong relationship with the same charge generation mechanism observed in photodetectors. The above observations indirectly confirm that the effects we are measuring are the consequence of light-induced surface charge changes. Figure 2d thus demonstrates that light illumination can indeed increase the generated osmotic power.

Ionic current measurements also show that the 643nm laser is more efficient for power generation than 475nm one, even though lower intensities were used. This fact is consistent with our PL measurements and the literature[20], and also originates from charge generation mechanism. As already mentioned, this is due to more efficient optical generation of charges with 643nm excitation since the energy of incident photons is closer to the excitonic resonance.

**Surface vs. bulk conductances.** We are intrigued by the fact that the osmotic voltages are higher in the larger nanopore (Figure 2c, 3nm and 10nm pore), even though the ion selectivity should be smaller due to a smaller EDL overlap. Experimental artefacts are unlikely since this behavior has been observed in other studies involving ultra-thin nanopores[4,5], though the origin of the effect was not resolved. To further investigate this counterintuitive behavior, we performed finite element simulations for different pore sizes. Consistent with decreasing EDL overlap, we always observed a decrease of the osmotic voltage with larger pores, independently of the surface charge assigned to the membrane (Figure S4a-d).

We suspect that small nanopores do not follow bulk conductance rules, since deviations in conductance from bulk predictions have been observed in nanopores with high aspect ratios L/D<1, where D is the pore diameter and L is the membrane thickness[22]. As a result, small nanopores might sense a modified chemical potential difference. In Supplemental Data, we derive a framework based on the Dukhin formalism that qualitatively explains the observed effect. The result shows that two competing mechanisms (**Figure** 4a) determine the ionic current through a nanopore of diameter D: the surface conductance and the bulk conductance. The importance of each contribution can be estimated through the Dukhin number: $Du = \frac{4 \cdot l_{Du}}{D}$, where $l_{Du}$ is the Dukhin length approximated by $l_{Du} \approx \frac{\sigma}{2 \cdot c_s \cdot e}$ with $c_s$ the bulk ion concentration and $e$ the elementary charge. The Dukhin number of the 3nm pore is 3.5 suggesting a large surface conductance contribution, whereas the 10nm pore has $Du=1$ (equal contributions). The classical Goldman-Hodgkin-Katz (GHK) equation[23] can estimate the osmotic potential when bulk conductance dominates, but overestimates the potential when the surface conductance becomes important. In this case, the ion concentration in the vicinity of the membrane plays a larger role than the bulk concentration (**Figure** 4b). We qualitatively estimated the osmotic potential for any pore size by weighing the GHK equation with $Du$ (**Figure** 4c, solid blue line):

$$E_{total} = \frac{RT}{F} [R_b \cdot ln(\frac{P \cdot c_{cis} + c_{trans}}{P \cdot c_{trans} + c_{cis}}) + R_s \cdot ln(\frac{P \cdot c_{K^+}^{cis} + c_{Cl^-}^{trans}}{P \cdot c_{K^+}^{trans} + c_{Cl^-}^{cis}})] \tag{1}$$

, where $P$ is the permeability ratio, $R_b$ and $R_s$ are the contribution ratios of the bulk conductance and surface conductance that satisfy $R_s + R_b = 1$ and are estimated using the Dukhin number: $R_b = \frac{1}{1+Du}$ and $R_s = \frac{Du}{1+Du}$. The ion concentration close to the charged membrane is denoted as: $c_{K^+}^{trans}$, $c_{Cl^-}^{trans}$, $c_{Cl^-}^{cis}$ and $c_{K^+}^{cis}$. This effectively scales the osmotic potential with the surface and bulk contributions. The trend observed in the experiments is qualitatively reproduced by this modified GHK equation. Strikingly, the highest osmotic potential is found for a pore size of ~9nm which is similar to the values observed experimentally (Feng et al., Figure 3d)[4].

**Consequence for light boosting.** The surface conduction highly dominates the smaller 3nm nanopore. This decreases the effective concentration gradient sensed by the nanopore. As a consequence, the osmotic potential is lower than what would be expected from bulk predictions. The larger the nanopore becomes, the less it is dominated by surface conduction, emphasizing the bulk gradient and therefore increasing the effective chemical potential difference. We have shown that light can double the osmotic power in small nanopores and increase the power output of larger pores by up to 20%. Since the power is a product of osmotic voltage and osmotic current it can be affected by either of the osmotic potential, or the osmotic current.

During light-exposure of the pore two changes are occurring simultaneously: First, the surface charge at the pore rim increases, leading to an enhanced ion selectivity (by enhancing repulsion of anions) and therefore larger osmotic voltage. Second, the surface charge of the membrane increases, which in turn enhances the surface conductance and therefore increases the osmotic current. The first effect dominates in small pores, while the second effect dominates in bigger pores. This can be seen by the astonishing osmotic potential increase in small nanopores, while in larger pores this difference is less pronounced and results only in a small improvement of the ion selectivity. However, even though the effect on the ion selectivity is small in the large pore, the enhanced surface conductance during light illumination (due to large $l_{Du}$) increases the ionic current substantially (Figure 2b). This effect is becoming increasingly important as the pore size increases and helps to maintain good power generation in large nanopores. Along the same line, recent work has analytically shown that strong surface conductance generates a dynamic ion selectivity, which is defined through Dukhin, rather than EDL, overlap and is responsible for the maintenance of good ion selectivity in larger pores[24].

**Outlook.** To increase the osmotic current, arrays of nanopores will have to be fabricated on a large-scale. We suggest achieving this through a non-specific large scale defect creation method, such as oxygen plasma[25] or ion bombardment[26]. This will produce a large pore size distribution, where small and large nanopores are present on the membrane at the same time. Light induced charge generation will render both nanopores more efficient. Increasing the reactivity of $MoS_2$ to light could provide a way of further boosting the permeability ratio in small nanopores, but equally important, help maintaining a good selectivity and strong surface current in larger nanopores. Self-assembled gold nanoparticles or plasmonic nanoshells could be used to induce a surface plasmon resonance and thus increase the charge creation[27,28]. Aluminium nanocavities can improve the absorbance of the light to nearly 70%[29], but would increase the membrane thickness. A better option might be to use a single-layer of hexagonal boron nitride (hBN) for that purpose[30].

In a real-world application, one could imagine boosting the osmotic power generation during daytime by directing sunlight onto the membranes. Typical irradiance levels of sunlight are about 100 mW/cm$^2$ at sea level making it easy to achieve here reported power densities by focusing the sun's light using standard concentrators used in solar cells[31].

To our knowledge, this is the first observation that laser light can modulate the ion selectivity of $MoS_2$ nanopores through a photogating effect and thus increase the osmotic power generated through reverse electrodialysis. Furthermore, we developed a novel formalism to qualitatively explain the data observed in thin nanopores.

## Experimental Procedures
### Nanopore Fabrication
The fabrication of the devices has been previously described[4,14]. In summary, we prepared silicon nitride membranes by using conventional photolithography and anisotropic etching of silicon in KOH. We then used e-beam lithography and reactive ion etching with a mixture of $CHF_3$ and $SF_6$ gases to etch through silicon and silicon dioxide to create 50nm sized holes in the nitride membrane.

Single crystal $MoS_2$ was grown using chemical vapor deposition on sapphire substrates. We then spin-coat PMMA on this substrate and do a water-based transfer to the target substrate. The devices are then soaked in Acetone and baked overnight at 400°C with a flow of argon(100sscm) and hydrogen (10sscm) in order to remove the majority of PMMA residues.

The devices are then placed into a transmission electron microscope (FEI Talos, Hillsboro, Oregon, USA) at an acceleration voltage of 80kV to assess the integrity and cleanliness of the sample. If the sample passes the quality control, a nanopore of the desired size is drilled by focusing the electron beam to the smallest possible spot.

The devices are then mounted into a custom made PMMA flow-cell using two rubber O-rings to ensure a good seal. The two chambers are subsequently flushed with an ethanol-DI water mixture (50/50) in order to properly wet the nanopore. The input channels are then sealed off by a cover slip to create a flat water-glass interface. This avoids light scattering and greatly improves the alignment precision.

The flow cell is then placed on a xy-stage in a custom-made Faraday cage and aligned to the laser beam. The laser electronics (iBeam smart, Toptica Photonics, Gräfelfing, Germany) are placed outside the metal cage to avoid noise increases. The beam is inserted into the cage through a hole in the box. On the opposite side of the flow cell a small USB microscope is placed to observe the light transmitted through the nitride membrane. The flow-cell can then be carefully aligned using the xy-stage. The point where the most light penetrates the nitride membrane was considered to be the ideal position.

An Axopatch 200B in combination with a custom made LabView program is used to measure and record the current through the nanopore.

Pore sizes were checked by comparing current-voltage characteristics in symmetrical salt concentration, i.e., 1M/1M KCl at the start and the end of the experiments. All pore sizes were estimated by substituting the conductance in the well-established conductance model[32]: $G = \sigma[\frac{4l}{\pi d^2} + \frac{1}{d}]^{-1}$, where $G$ is the conductance, $d$ the pore diameter, $l$ the pore length (set to 1nm for MoS$_2$) and $\sigma$ the ionic conductivity.

For all measurements, we used the potassium chloride (K$^+$/Cl$^-$) ion pair due to their nearly identical bulk mobilities. All our buffers have been adjusted to a pH of 7.4, except for the 1mM dilution, which was adjusted, but unbuffered.

All laser powers reported are the values that have been entered by the user in the software (iTopas, Toptica Photonics, Gräfelfing, Germany). A table of the theoretical vs. measured output power can be found in Table S1. Roughly, the measured power was about 90% (643nm) and 98% (475nm) of the entered value. In practice, the power reaching the membrane is less, due to scattering at the coverslip and in the PMMA flow-cell.

The laser fluxes were calculated by dividing the corrected output power P$_c$ by the approximate area that have been measured using a slit scanning beam profiler (BP104-VIS, Thorlabs, Newton, New Jersey, United States). The correction factor was determined by measuring the laser power using a FieldmaxII-TOP (Coherent, Santa Clara, United States) and reported in Table S1. The loss is roughly 10% for 643nm and 2% for 475nm. We can thus calculate the fluxes as follows: $I = (1 - loss) \cdot P_0 \cdot \frac{1}{\pi \cdot X_{4-\sigma} \cdot Y_{4-\sigma}}$, $I_{643nm} = P_0 \cdot 15.05[\frac{W}{cm^2}]$,
$$I_{475nm} = P_0 \cdot 27.25[\frac{W}{cm^2}]$$

**Optical measurements**
Micro photoluminescence (µ-PL) spectra were performed in air at room temperature using the laser light focused to the diffraction limit with a beam size of about 1 µm. The incident power was 90 µW for both excitation wavelengths. The emitted light was acquired using a spectrometer (Andor) and the laser line was removed with a long-pass 488-nm (650-nm) edge filter in the case of 488 nm (647 nm) excitation. The presented µ-PL maps were obtained by scanning the sample using the nano-positioning stage (Mad City Labs Nano-Drive). Bright field image was acquired by a CCD camera (Andor Ixon).

**Data analysis**

The IV traces were recorded at a reduced sampling rate 6'250Hz with a lowpass filter set to 1kHZ. The voltage polarity switched when the voltage ramps to avoid charging of the membrane, i.e. 0, 100mV, -100mV, 200mV, -200mV. Each voltage step lasts for 10 seconds. The traces are then segmented through detecting the current transient when the voltage was switched. Each extracted part is fit to a simple exponential function $I(t) = I_0 \cdot e^{-b \cdot x} + I_{stable}$. The considered current value corresponds to the stable current at infinite time. The standard deviation of this fit µ$_{exp\text{-}fit}$ is extracted by taking the square root of the covariance matrix. We then perform a linear fit of the obtained current-voltage values. By using York's method we can propagate the previously calculated errors correctly through the linear fit [33]. We can now extract the parameters of the current-voltage relationship $I(V) = G \cdot V + I_0$, where $G$ is the conductance. The measured osmotic current is defined as $I_0$, the measured osmotic potential is $\frac{I_0}{G}$. The redox potential of the Ag/AgCl electrodes are the removed from the measured osmotic potential to yield the effective voltage $V_{osm} = \frac{I_0}{G} - E_{redox}$. The electrode potential of the Ag/AgCl electrodes have been measured for the buffers used in the experiments using a reference electrode. Table S2 reports the measured and theoretical voltages. The deviations to the theoretical values most probably come from the uncertainty of preparation of such dilute solutions.

The effective osmotic current is then calculated as $I_{osm} = V_{osm} \cdot G$. The osmotic power is defined as: $P_{osm} = V_{osm}^2 \cdot G$
The errors are propagated using the following formula:
$$s_f = \sqrt{(\frac{\partial f}{\partial x})^2 s_x^2 + (\frac{\partial f}{\partial y})^2 s_y^2 + (\frac{\partial f}{\partial z})^2 s_z^2 + \cdots}$$
The error on the redox potential is calculated by assuming a 5% error in the salt concentration of the two reservoirs c$_{min}$ and c$_{max}$ using the following equation:
$$\mu_{E_{redox}} = \frac{RT}{F} \cdot \sqrt{(\frac{1}{c_{max}^2} \cdot \mu_{c_{max}}^2 + \frac{1}{c_{min}^2} \cdot \mu_{c_{min}}^2)}$$
Similarly:
$$\mu_{V_{osm}} = \sqrt{\frac{1}{G^2}\mu_{I_0}^2 + \frac{I_0^2}{G^4}\mu_G^2 + \mu_{redox}^2},$$
$$\mu_{I_{osm}} = \sqrt{\mu_{I_0}^2 + \mu_G^2 E_{redox}^2 + \mu_{redox}^2 G^2} \text{ and}$$
$$\mu_{P_{osm}} = \sqrt{(E_{redox} - \frac{I_0^2}{G^2})^2 \cdot \mu_G^2 + (\frac{2I_0}{G} - 2E_{redox})^2 \cdot \mu_{I_0}^2 + (2GE_{redox} - 2I_0)^2 \cdot \mu_{E_{redox}}^2}.$$

## COMSOL Numerical Modeling –Heat

COMSOL 5.3a was used for all finite element simulations. The COMSOL model was built by keeping all dimensions as close to the reality as possible. The silicon chip was defined as 380um thick and 5x5mm square. The backside contained a 520um large, pyramidical opening penetrating through the chip and leaving a 52µm large membrane area on the frontside. The silicon chip is encompassed by a 10mm cubic PMMA flowcell, which has two 1mm large channels that provide the liquid contact to the silicon chip. The *heat transfer in solids* module is used. Initial temperatures were set at 20°C. We then apply the power from the laser in two steps by using the *Deposited Beam Power* module. 1st, we apply a Gaussian distributed beam (centered in the middle of the membrane, standard deviation 250um) of power of 70% of the laser power (simulating the silicon absorbance) to the area outside of the membrane. 2nd, we apply a second beam (same center and standard deviation) of power of 10% of the laser power (nitride absorbance) to the membrane only. *Heat transfer in Solids and Fluids* modules are used to model transfer of heat along the silicon chip, the PMMA and the liquid. Furthermore, the *diffuse surface* module is used to take into account the radiative heat source. The finite element simulation is then run for different laser power. As a boundary condition for the temperature, the edge of the PMMA flow-cell was set to be at 20°C.

## COMSOL Numerical Modeling –Nanopore

We simulated the stationary ion distribution around a monolayer $MoS_2$ membrane by solving the Poisson-Nernst-Planck equations given by:

$$\nabla^2 \phi = \frac{eF}{\epsilon}(K^+ - Cl^-) \tag{1}$$

$$\nabla \cdot J_{K^+} = \nabla \cdot \left[ -F D_{K^+} \nabla K^+ - \frac{F^2 D_{K^+}}{RT} K^+ \nabla \phi \right] = 0 \tag{2}$$

$$\nabla \cdot J_{Cl^-} = \nabla \cdot \left[ -F D_{Cl^-} \nabla Cl^- - \frac{F^2 D_{Cl^-}}{RT} Cl^- \nabla \phi \right] = 0 \tag{3}$$

Where ϕ is the electrostatic potential at every point in space. This potential along with the current densities of the respective ionic species ($J_i$) is dependent on the concentration of the ionic species (i = $K^+/Cl^-$), the respective diffusion coefficients ($D_i$) and the temperature *T*. *F* is Faraday's constant while *R* is the universal gas constant and ε is the relative permittivity. Equation (1) refers to the self consistent Poisson equation that computes the electrostatic potential in the presence of the solute ions while eqs 2 and 3 refer to the Nernst-Planck equation describing the electrodiffusion in terms of concentration of the ionic species across the membrane. Since we are interested in the steady state distributions of the ions in the vicinity of the nanopore, the current continuity equations are set to 0. A modified version of the above equations was used to explore the ionic distribution around nanopores in salt concentration gradients by Xie et. al.[34]

The equations 1-3 were solved simultaneously using COMSOL 5.3a using The *Transport of Diluted Systems* module coupled to the *Electrostatics* module. We fixed the salt gradient to be 1mM on the trans side and 100mM on the cis side, by defining the corresponding concentration as the boundary conditions on the top and bottom walls. Similarly, the boundary conditions on the electric field were set by defining the upper wall of the cis side as a ground whereas the bottom wall of the trans side was set to the applied voltage. The surface charge of the pore rim was set to 2 times the surface charge of the top and bottom surface to address the higher reactivity of the pore rim. We then sweep through different pore sizes, surface charges and applied voltages. By measuring the net ionic current through the nanopore for different applied voltages, we can simulate the current voltage relationships.

We can then use the same method as applied to the experimental data to extract the osmotic current and potential. A similar setup was utilized by Rollings et. al. [5] to investigate ion selectivity in graphene nanopores.

The continuum PNP model described above is quite useful to investigate electrodiffusion of charged species in ion channels, however there are a few known shortcomings. These include the neglect of finite volume effect of the ionic particles and correlation effects (such as ion-ion interactions and steric effects). While there have been a few known corrections and modifications to the PNP theory, they have not been considered in our model. Alternately, many other ab-initio methods can also be used to investigate electrodiffusion in narrowly confined ion channels such as Molecular Dynamics (MD), Brownian Dynamics (BD) and Monte-Carlo simulations. Each of these methods offer their own unique advantages to analyze ionic flows at varying length scales which we are yet to explore.

### Acknowledgements
This work was financially supported by the Swiss National Science Foundation (SNSF) Consolidator grant (BIONIC BSCGI0_157802), CCMX project ("Large Area Growth of 2D Materials for device integration") and EPFL's ENABLE program. We thank the Centre Interdisciplinaire de Microscopie Electronique (CIME) at the École Polytechnique fédérale de Lausanne (EPFL) for access to electron microscopes. Device fabrication was partially carried out at the EPFL Center for Micro/Nanotechnology (CMi). We would like to thank Dr. Jochem Deen, Dr. Andrew Laszlo and Mukeshchand Thakur for helpful discussion on the manuscript. Furthermore, we would like to thank Michal Macha, Dr. Dumitru Dumcenco, Dr. Huanyao Cun, Yanfei Zhao, Mustafa Fadlelmula for providing $MoS_2$ material.

### Author contributions
M.G. and A.R. conceived the idea and designed all experiments. M.G., M.L. and A.R. wrote the manuscript. M.G. and M.L. performed the measurements. M.G. performed the data analysis. M.G. and M.L. fabricated the nanopore devices. D.U. performed PL measurements under A.K.'s supervision. M.G. and A.S. built the COMSOL models. All authors provided constructive comments to the manuscript.

# Figures

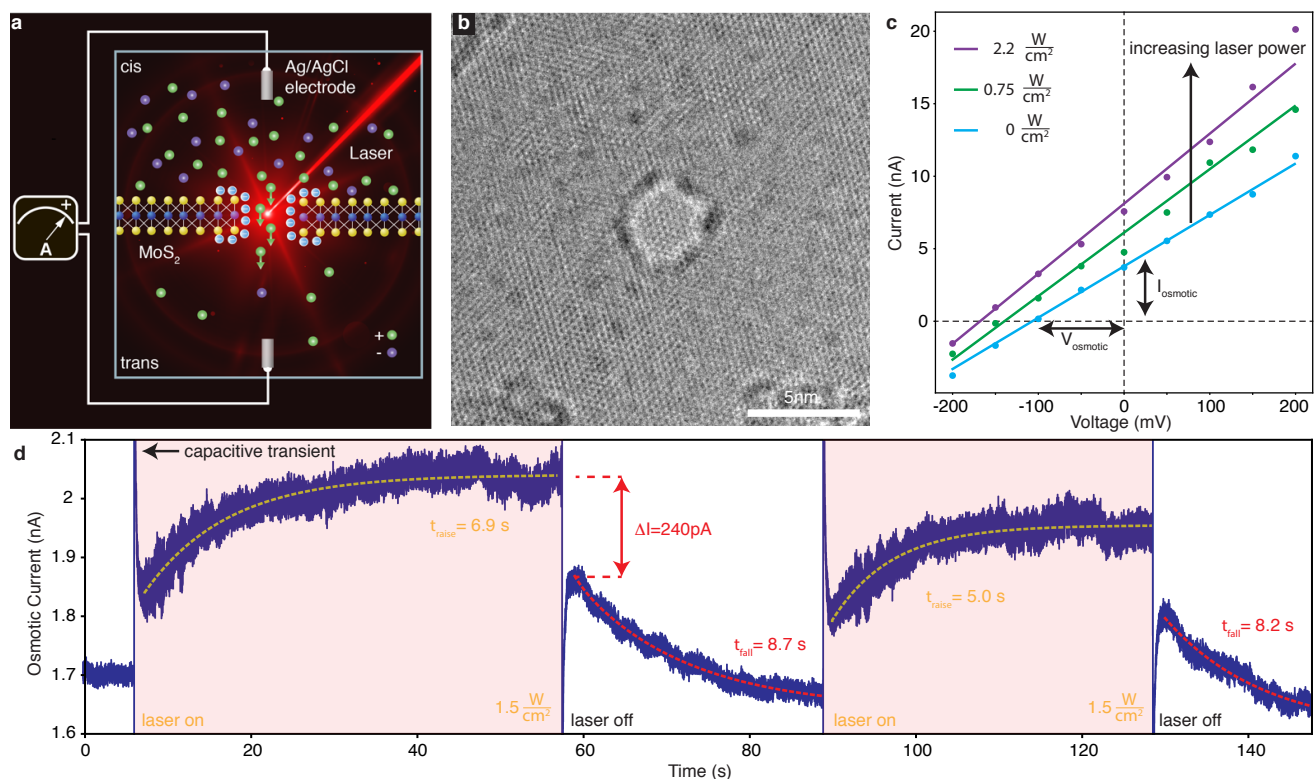

***Figure* 1: Experimental Overview. a**, Schematics of the experimental set-up: Laser light is used to photogate the $MoS_2$ and thus modulate the ion selectivity through the nanopore. **b**, A typical nanopore drilled through an $MoS_2$ membrane by focusing the beam of an transmission electron microscope using an acceleration voltage of 80kV[14]. **c**, IV characteristics at a concentration gradient of 1000 using different laser intensities. **d**, Time trace of the osmotic current while switching the laser on and off (643nm, 1.5 W/cm²). The half-life values have been extracted by fitting to a single-exponential function

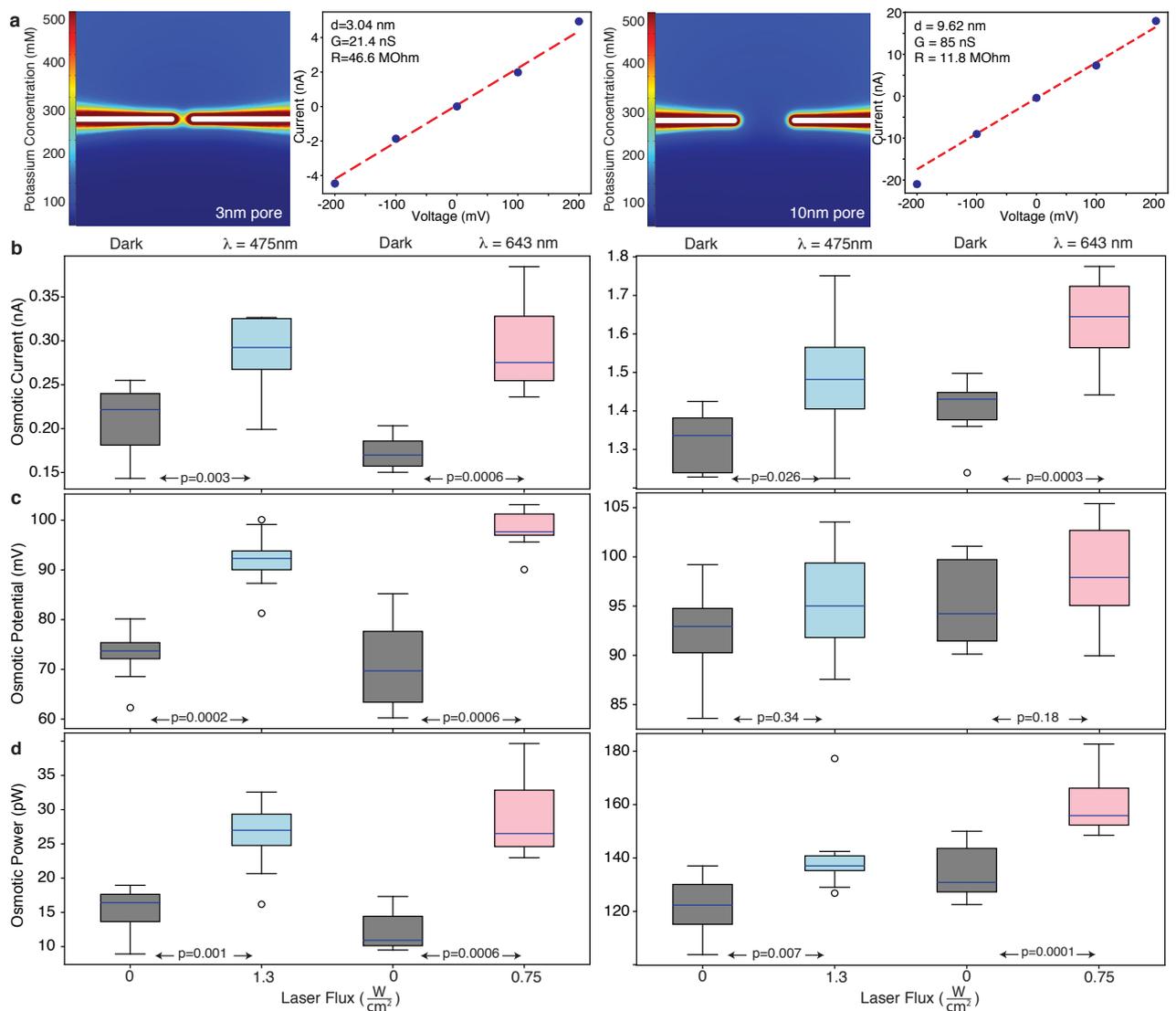

*Figure 2: **Effect of laser light on the Osmotic Energy Conversion for a 3nm (left side) and 10nm (right side) pore. a**, The potassium concentration distribution calculated through a finite element simulation for a 3nm (left) and a 10nm (right) pore. The osmotic current (**b**), voltage (**c**) and power (**d**) generated by a 3nm (left) and 10nm (right) MoS$_2$ nanopore as a function of laser (grey: laser off, color: laser on) and wavelength (red: 643nm, blue: 475nm). The blue line indicates the median values, the upper and lower ends of the box denote the upper and lower quartile, whereas the whiskers encompass the rest of the data points. n=10 for all the measurement groups for the 3nm pore, n=11 for the 643nm laser for pore size 10nm and dark state, n=12 for 475nm laser and its dark state. P-values reported are calculated using the Wilcoxon rank-sum test and test for the alternative hypothesis that samples from the dark state distribution are more likely to be smaller than samples from the illuminated distribution. All p-values are highly significant (p<0.05), except for osmotic potential values for the 10nm pore. Mean values for each distribution are reported in Table S3.*

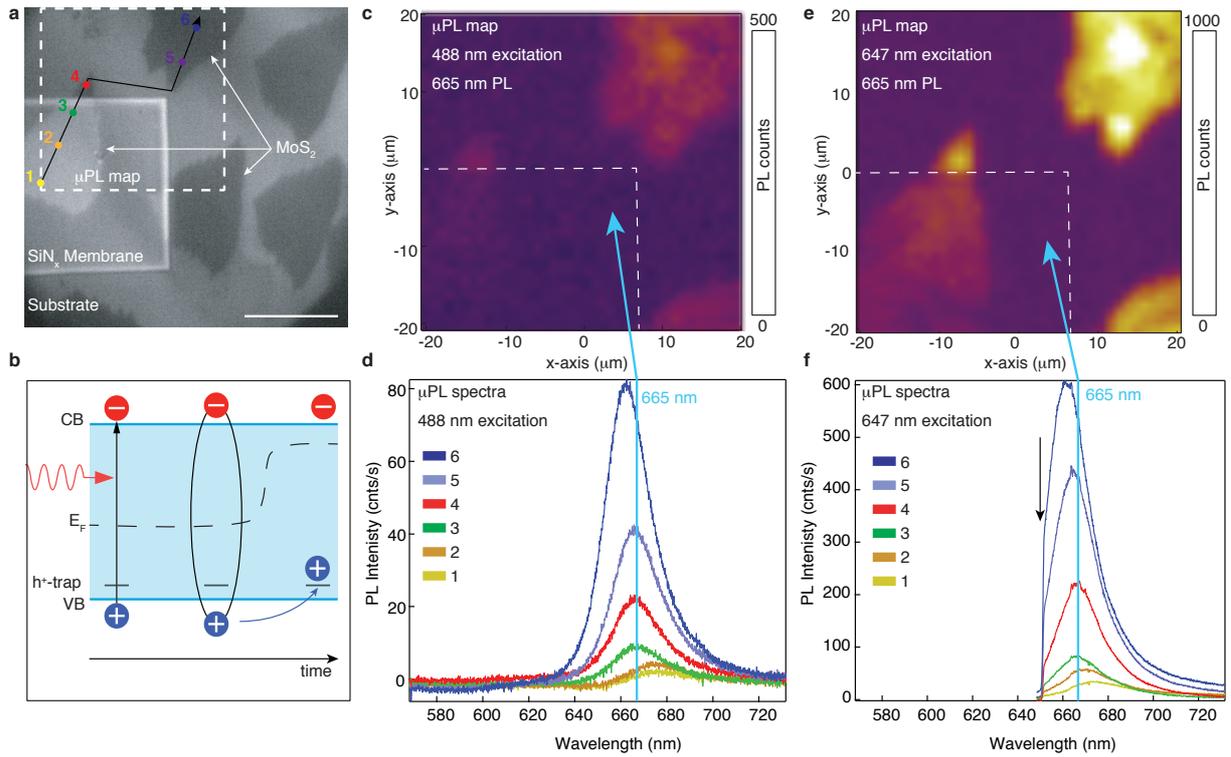

***Figure 3: Photoluminescence Measurements. a,*** *Optical image of the device area used for the photoluminescence measurements. The dashed rectangle represents the position of the µPL maps in (**c**) and (**d**). Color points show position for µPL spectra in (**d**) and (**f**). Scale bar is 20 µm.* ***b,*** *schematic depiction of the photogating effect in a single-layer MoS$_2$, when trapped photoexcited holes shifts Fermi level closer to the conduction band (CB).* ***c*** *and* ***e*** *µPL maps acquired at high energy (488 nm) and near-resonant (647 nm) excitations correspondingly. Both maps demonstrate PL intensity distribution at 665 nm emission wavelength. Resonance excitation results in an 8-fold PL intensity enhancement. The dashed rectangles shows edges of the membrane.* ***d*** *and* ***f*** *Corresponding µPL spectra. Black arrow represents near-resonant excitation at 647 nm wavelength.*

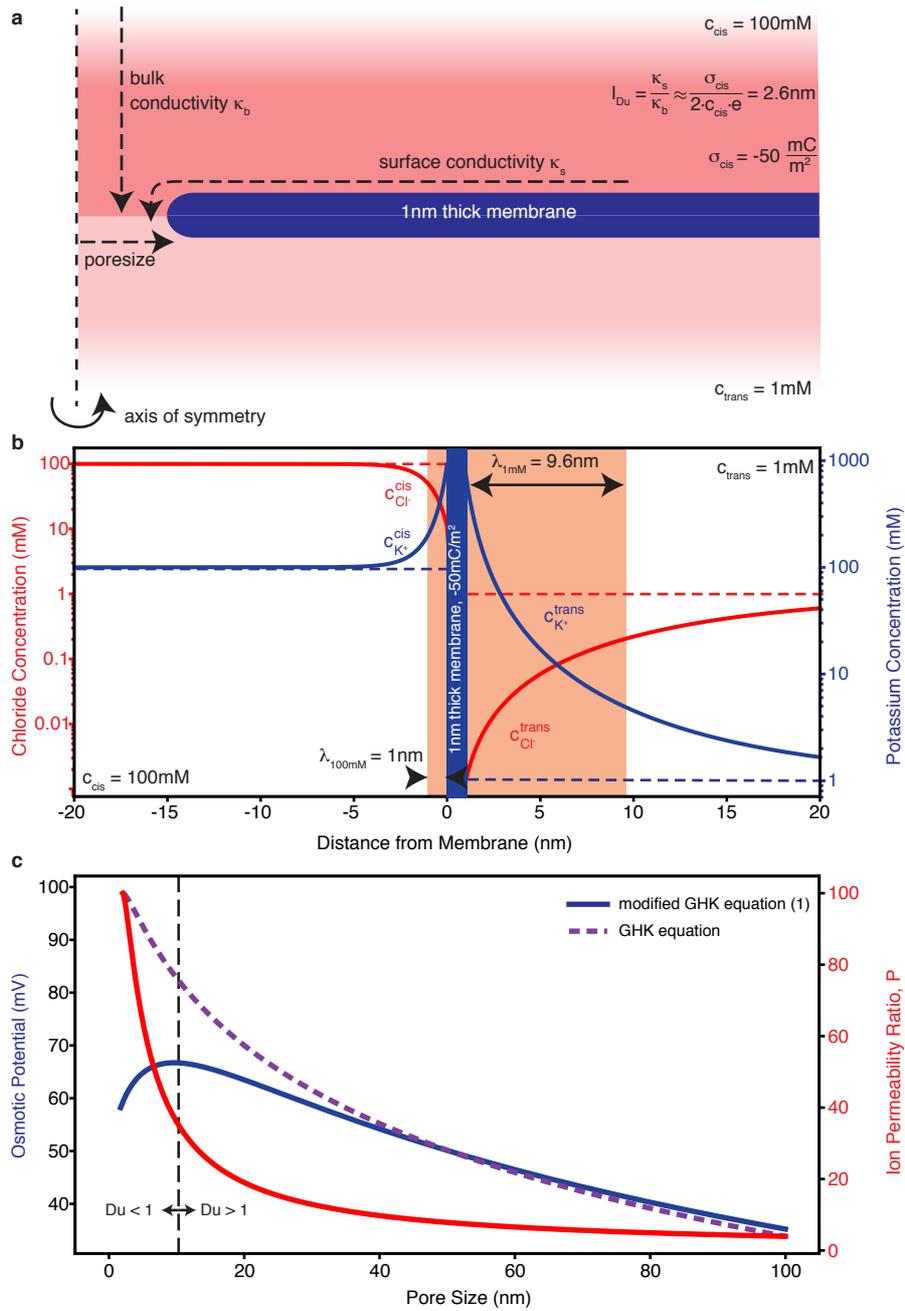

*Figure 4: **Surface Conduction contribution. a,** Illustration of bulk and surface conductance contributions. In high aspect-ratio nanopores, the surface conductivity is an important factor, quantified by the ratio of surface-to-bulk conductivities and denoted by the Dukhin length $l_{Du}$. This value can be estimated through the ionic strength of the bulk c and the surface charge $\sigma$ through $l_{Du} \approx \frac{\sigma}{2 \cdot c_s \cdot e}$. **b,** Simulated potassium (blue) and chloride (red) concentration distribution from a charged wall with surface charge $\sigma_{cis}$=-50mC/m². The bulk concentration levels of potassium and chloride $c_{cis}$ and $c_{trans}$ are indicated with the blue and red dashed lines. The red box shows the EDL length over which ions have been averaged for the calculation of the effective ion concentration in the case of a surface conductance dominated system. **c,** The reversal potential calculated from the modified GHK model (equation 1) is denoted with a solid blue line and plotted versus pore size (D>2nm). This model scales the two competing effects through prefactors weighing the bulk contribution and the surface contribution. The values of the prefactors are a function of the pore size D and are calculated through the Dukhin number Du. The dashed blue line shows the non-modified GHK equation. The Ion permeability assumed for the calculations is plotted in red on the secondary y-axis and is estimated from a geometrical consideration of EDL overlap (see Supplemental Information). The vertical black line at around 10nm pore size shows where the Dukhin number passes 1, i.e. at which pore size the system switches from a surface conductance dominated system to a bulk conductance dominated system.*

# Supplemental Information

## Light Enhanced Blue Energy Generation using MoS$_2$ Nanopores

Michael Graf[1], Martina Lihter[1], Dmitrii Unuchek[2], Aditya Sarathy[3], Jean-Pierre Leburton[3], Andras Kis[2], Aleksandra Radenovic[1]

[1]Laboratory of Nanoscale Biology, Institute of Bioengineering, School of Engineering, EPFL, 1015 Lausanne, Switzerland
[2]Laboratory of Nanoscale Electronics and Structures, Institute of Electrical Engineering and Institute of Materials Science and Engineering, School of Engineering, EPFL, 1015 Lausanne, Switzerland
[3]Department of Electrical and Computer Engineering, University of Illinois at Urbana-Champaign, USA

## Overview of Figures and Tables

## Detailed Discussion on how heat influences our system

First, we set out to estimate the heating effect of the laser power when it penetrates through water. We used the following equation derived by [1] to estimate the steady state change of temperature ΔT:
$\Delta T = \frac{\alpha}{2\pi \cdot C} \cdot [ln(\frac{2\pi \cdot R}{\lambda}) - 1] \cdot P$, where α is the absorption coefficient (*0.0114 for 475nm and 0.322 for 643nm*[2]), λ the wavelength of the laser used, P the laser power (150mW, full theoretically possible power), C the thermal conductivity of water (0.60 Wm$^{-1}$K$^{-1}$) and R the distance to the surface of the flow cell. For both wavelengths, the temperature increases are below 0.01°C, which means that we can neglect any heating effects of the water.

Since we are working with a non-focused laser of about 1mm spot size (Figure S1) we irradiate not only the silicon nitride membrane but also some of the silicon that is found outside the membrane. To estimate the heat produced by this system we simulated the geometry in COMSOL. We designed a 50µm squared, 20nm thick nitride membrane in a 5mm dice of silicon. Water is placed in a circular fashion on top and on the bottom of the chip. A 10x10mm PMMA block is placed around the system and the outer boundaries of this block have been set to room temperature. An illustration of the simulated geometry can be found in Figure S5a. The silicon nitride window and the silicon are treated separately using two instances of the *deposited beam power* module. Since the silicon nitride window is transparent, we assume an absorbance value of only 10% as previously calculated[3]. On the other hand, the absorbance of the silicon part was set to 70%, since 30% of the light gets reflected on its polished surface[4] (Figure S5b). The spot size to calculate the power flux was set to 1mm, whereas the deposited Gaussian beam profile was set to a standard deviation of 250μm to correspond to the commonly used 4-σ value of beam width. The resulting temperature profile in the case of a 1.5mW/cm$^2$ intensity can be found in Figure S5c. Figure S5d shows the temperature distribution along the z-axis of the nanopore chip for different laser powers. A small peak is found at the nitride membrane. In general, the highest possible temperature increase at its peak value is about 4°C. These values are absolute upper limits, since we do not take into account any losses happening due to scattering on the PMMA flow cell and intensity losses when the light penetrates the glass slide covering the flow cell. We can expect that the real temperature increases to be substantially lower than the values estimated here.

In this part we estimate and rule out the influence of temperature on the observed results. Here, we ignore any influence of the surface charge concentrate on the temperature dependence of the KCl conductivity. Electrolytic conductivity (σ) values of KCl (concentrations ranging from 10mM to 1M) in a temperature (T) range of 0 to 50°C were extracted from the NIST standards for electrolytic conductivity[5]. Since the conductivity-temperature relationship is linear in this range, we extracted parameters $a$ and $b$ from: $\sigma = a \cdot T + b$. We can then extract the temperature corresponding to a certain nanopore conductance using: $T = \frac{\sigma - b}{a}$, where $\sigma = G \cdot [\frac{4l}{\pi d^2} + \frac{1}{d}]^{-1}$ [6]. In these calculations, we assume that the pore diameter and length does not change. We used a suspended MoS$_2$ layer in a 50nm nitride hole. In this configuration, the MoS$_2$ layer reaches until the nitride edge producing a pore of diameter 50nm and a thickness of 21nm (20nm nitride + ~1nm MoS$_2$). The advantage of this configuration is its higher temporal stability. In Figure S6 we measured the conductance of this nanopore at different symmetric salt concentrations of 1M, 100mM and 10mM. If we apply the same temperature analysis, we observe that higher temperature differences are needed to explain the data at low salt concentration (up to 60°C). This is not consistent with the reasonable assumption that the laser heating is independent of the ion concentration. There must be another, concentration dependent variable at play: the surface charge.

Other than viscosity changes, heat also influences the chemical potential difference. We can express the osmotic voltage observed in Figure 2 in a simplified way using the reversal potential:
$V_{diff} = S(\Sigma)_{is} \frac{RT}{F} \cdot ln[\frac{c_{cis}}{c_{trans}}]$, where V$_{diff}$ is the measured osmotic potential, S(Σ)$_{is}$ the ion selectivity and the logarithmic expression the concentration gradient. To estimate the pure thermal effect, we assume that the laser is not influencing the ion selectivity, so we set S(Σ)$_{is}$ to a fixed value (selectivity in the dark state) and vary T in order to obtain the values measured. In the case of the smaller, 3nm pore, a temperature differences of 118°C and 73.9°C are needed to explain the change due to a 643nm and 475nm laser irradiation. For exactly the same laser conditions these values drop to 11.5°C and 6.5°C for the 10nm pore. Realistically, no difference should be seen between the two cases since an enlarged nanopore does not influence how the laser heats the system. Furthermore, the values obtained for the small nanopore are at least an order of magnitude away from anything we could expect from the previous estimation.

In order to estimate the influence of the temperature on the EDL we analytically calculate the thickness of the EDL for different temperature (Figure S7). The thickness of the EDL corresponds to the Debye length. With the Debye-Hückel approximation we can calculate the Debye length λ as[7]: $\frac{1}{\lambda} = \kappa = (\frac{e^2 \sum n_i^\infty z_i^2}{\varepsilon_0 \varepsilon_r k_B T})^{\frac{1}{2}}$, where κ is the Debye-Hückel parameter, $\varepsilon_r$ the relative permittivity of water, $\varepsilon_0$ the permittivity of vacuum, $k_B$ Boltzmann's constant, $T$ temperature, $n_i$ the bulk volume density, $z_i$ the valence (in the case of KCL: $\sum n_i^\infty z_i^2 = 2n^\infty$ ) and $e$ the electron charge. Since $\varepsilon_r$ depends on the temperature as well we can use an analytical approximation to calculate its value[8]. Since the EDL length actually decreases a few picometers per °C of increased temperature, we cannot expect to see any improvement of the pore ion selectivity due to temperature changes and EDL thickness.

The surface charge of MoS$_2$ in water is estimated through the following chemical equilibrium:
$$MoS_2 + H_2O \rightleftharpoons MoS_2 - OH^- + H^+$$

We can estimate the diffuse layer electrostatic potential Φ$_s$ (Zeta potential) of the surface by[9,10]:
$\Phi_s = \frac{K_B T}{e}(ln\frac{-\sigma}{e\Gamma + \sigma} + ln(10)(pK - pH))$, where σ is the surface charge and Γ the density of reactive sites. The pK of nanocrystalline MoS$_2$ has been measured to be around 3.1 [11]. We estimate the surface potential at different temperatures and pH using a pK value of 3.1 and a surface charge value of -50 mC/m$^2$. The surface potential changes with temperature, but the rate of this change is highly dependent on the pH as calculated in Figure S7c. For instance, at pH 7 the rate is just below 1 mV/K whereas at pH 4 the rate is only about 640 µV/K. Such an increase in surface potential might well improve the repulsion of cations and therefore increase the ion selectivity. Assuming we get a temperature increase of 10°C then we could expect a 10mV stronger surface potential. To put this value into context we can estimate the pH change needed to induce the same increase in surface potential. From Figure S7d we can see that the surface potential reduces 58mV per pH unit. In order to get a 10mV decrease, we would thus need a pH change of roughly 0.2, which is well within the error of our buffer system, especially at low dilutions.

Last but not least, the redox potential generated at the interface of the Ag/AgCl electrode with the ionic solution also depends on the temperature. Depending on the salt concentration the potential can increase between 200 and 600µV/K (Figure 7b). This is relevant if the whole system is heated and has to be considered if one wants to calculate the effective osmotic power generated by the membrane. In the case of laser irradiation, we can neglect any influences originating from the redox potential since we are not affecting the temperature that far away from the membrane.

# Derivation of the modified GHK equation

Deviations in conductance from bulk predictions have been observed in nanopores with high aspect ratios L/D<1[12] and were linked to a large contribution of the surface conductivity, described as the ratio between surface and bulk conductivity: $l_{Du} = \frac{\kappa_s}{\kappa_b}$, where $\kappa_s$ is the surface conductivity and $\kappa_b$ the bulk conductivity. This Dukhin length $l_{Du}$ can be approximated using the surface charge σ and the bulk ion concentration $c_s$ as: $l_{Du} \approx \frac{\sigma}{2 \cdot c_s \cdot e}$. The surface charge of the MoS$^2$ membrane was fixed to -50mC/m2 as previously determined[13]. The Dukhin lengths on both sides are then 2.6 and 26nm respectively (Figure 4a). Since this formalism has been developed for symmetrical salt concentrations, we set the effective Dukhin length to 2.6nm and thus provide the lower limit of the effect by underestimating the surface conduction effect. To further quantify the contributions of the surface and bulk conductances to the ionic current inside the pore we refer to the Dukhin number which is defined as: $Du = \frac{4 \cdot l_{Du}}{D}$, where D is the pore diameter[12]. The Dukhin number of the 3nm pore is 3.5 suggesting a large surface conductance contribution, whereas the 10nm has Du=1 (equal contributions). In order to calculate the osmotic voltage in the surface conductance dominated regime, we need to know the distribution of the ions in the vicinity of the membrane. Using FEM simulations, we estimate these ion concentrations as a function of distance to a charged membrane (σ=-50mC/m$^2$, Figure 4b). We then take the mean concentration from the wall (x=0) to a distance x=λ (where λ is the Debye length, λ=1nm for cis side and λ=9.6nm on trans side) to estimate the concentrations at the membrane surface: $c_{K^+}^{trans}$, $c_{Cl^-}^{trans}$, $c_{Cl^-}^{cis}$ and $c_{K^+}^{cis}$. The bulk concentrations of potassium and chloride are identical and simply denoted as $c_{cis}$ and $c_{trans}$. We rewrite the GHK equation to account for the surface conductance as well as the bulk conductance:

$$E_{total} = \frac{RT}{F}[R_b \cdot ln(\frac{P \cdot c_{cis} + c_{trans}}{P \cdot c_{trans} + c_{cis}}) + R_s \cdot ln(\frac{P \cdot c_{K^+}^{cis} + c_{Cl^-}^{trans}}{P \cdot c_{K^+}^{trans} + c_{Cl^-}^{cis}})]$$

Where P is the permeability ratio, $R_b$ and $R_s$ are the contribution ratios of the bulk conductance and surface conductance that satisfy $R_s + R_b = 1$ and are estimated using the Dukhin number: $R_b = \frac{1}{1+Du}$ and $R_s = \frac{Du}{1+Du}$, effectively scaling the reverse potential to surface and bulk contributions. We estimate the permeability values *P* by geometrically estimating the area affected by the EDL inside the pore:

$P(D) = P_{max} \frac{A_{ELD}}{A_{total}} = P_{max} \frac{\pi(\frac{D}{2})^2 - \pi(\frac{D}{2}-\lambda)^2}{\pi(\frac{D}{2})^2}$, which can be simplified to $P(D) = 4\lambda \cdot P_{max} \frac{D-\lambda}{D^2}$. The Debye length $\lambda$ was defined as 1nm inside the nanopore. The value of the maximal permeability ratio, $P_{max}$ was chosen to be 100 to best reflect the values obtained through FEM simulations (Figure S4a) and experimental studies[14].

## Figures

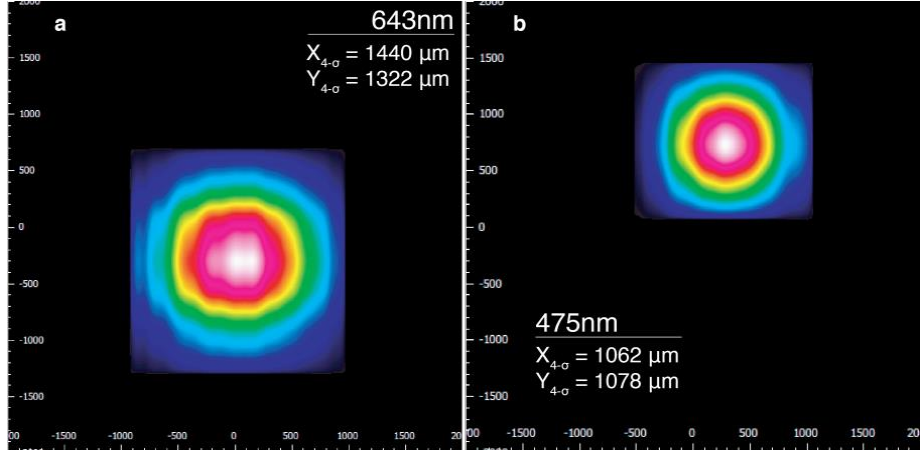

*Figure S1: **Laser Characteristics.** Measured laser spot sizes for laser the two diode lasers at 643nm (**a**) and 475nm (**b**).*

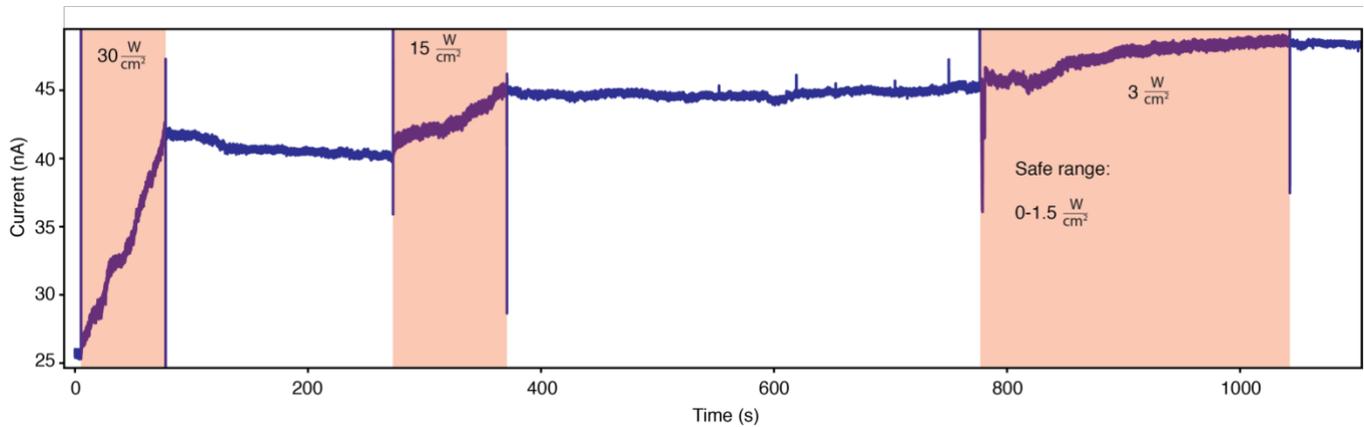

*Figure S2: **Nanopore Enlargement.** Enlargement of the nanopore for different photon fluxes for the 643nm laser*

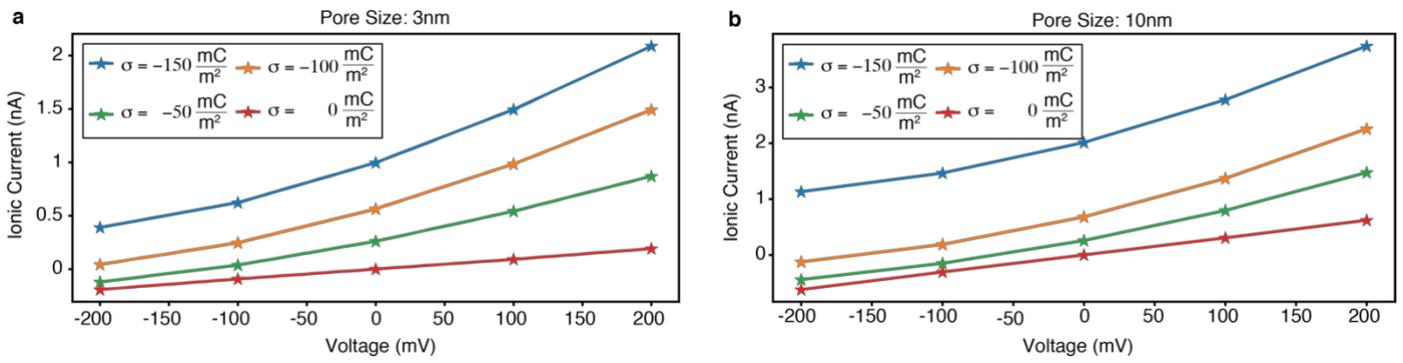

Figure S3: **IVs from FEM Simulations.** Simulated IV characteristics for the 3nm (**a**) and 10nm (**b**) pore. The experimentally observed upwards shift of the IVs is reproduced by this COMSOL model.

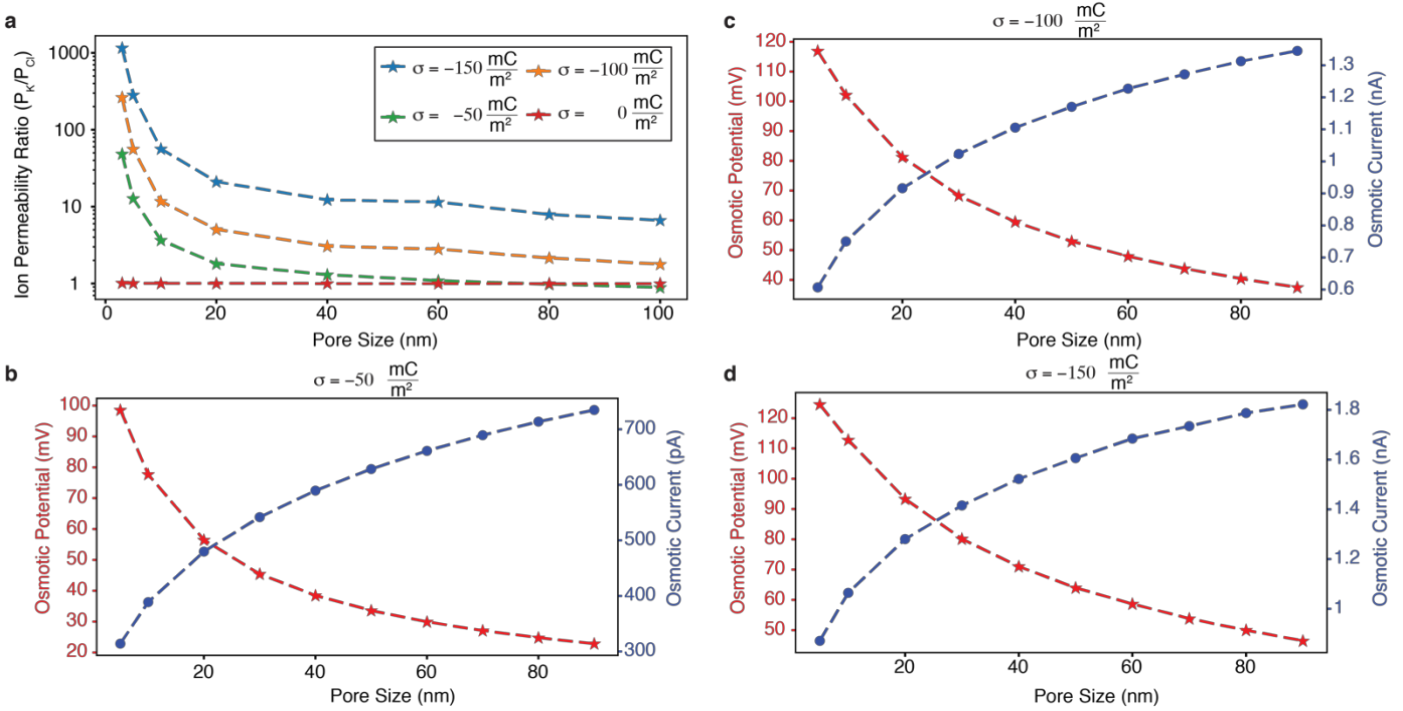

Figure S4: **FEM Simulations of permeability ratio, osmotic current and potential. a,** Simulated permeability ratio. **b-d,** Osmotic voltage and current for versus pore sizes for surface charges of -50 (**b**), -100 (**c**) and -150 mC/m² (**d**).

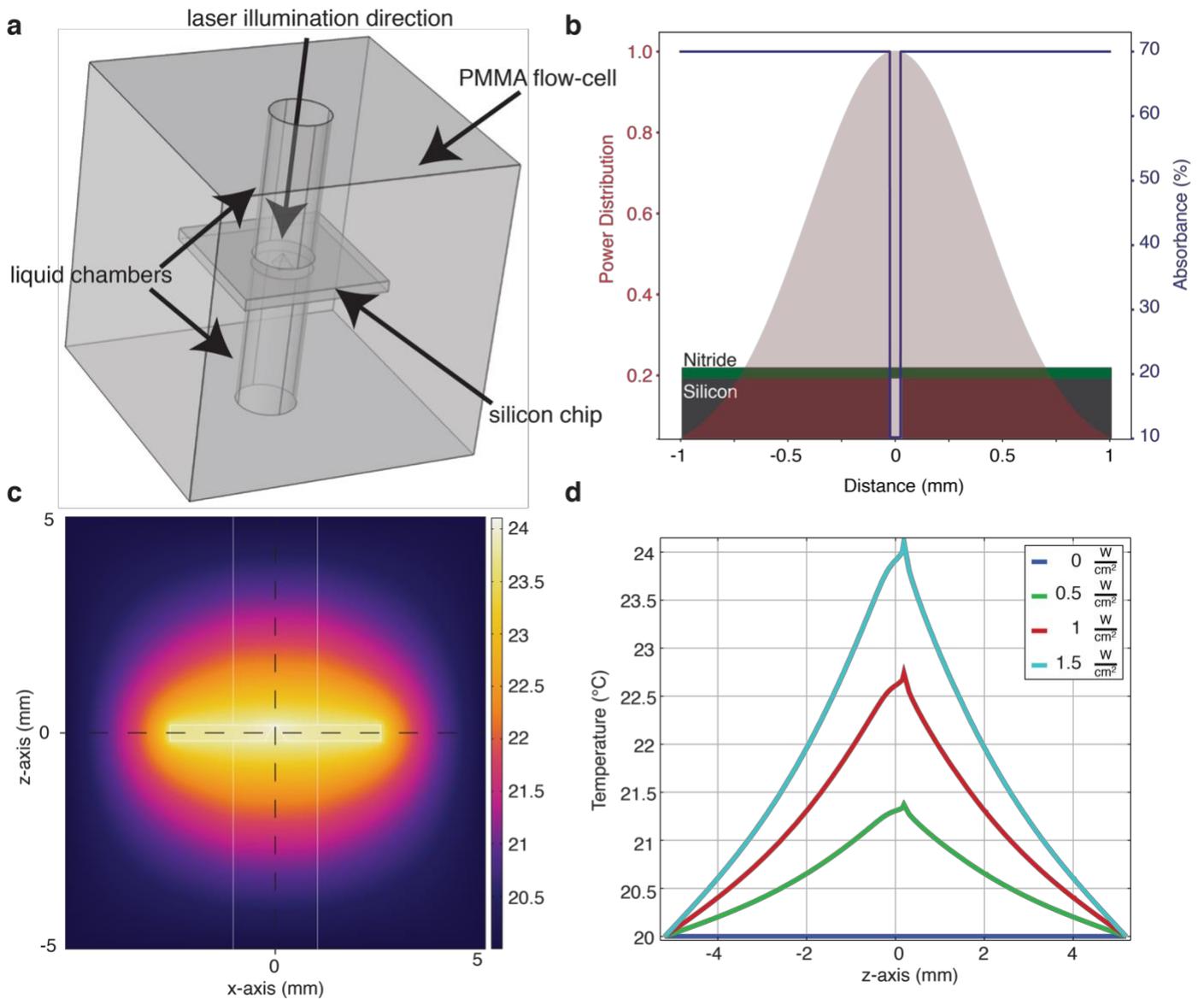

*Figure S5:* **COMSOL model used to estimate the heat generated by the laser light on the system. a,** *The geometry of the used model.* **b,** *Illustration of the power distribution irradiating the surface of the chip (red) and the different absorbance values over the surface.* **c,** *Heat map of the z-x axis of the system.* **d,** *The temperature profile along the z-axis and through the centre of the membrane.*

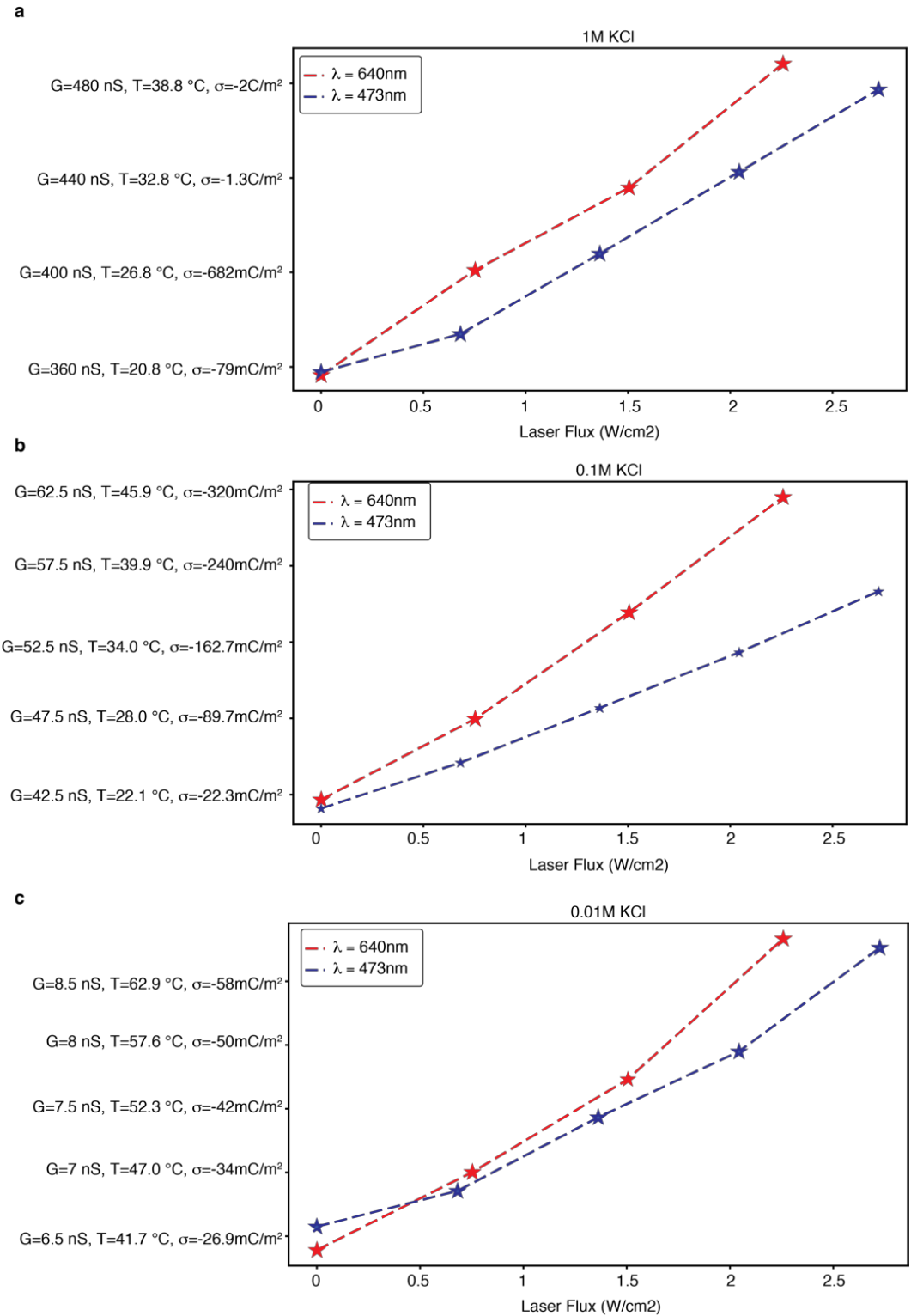

*Figure S6: **Laser Influence on the conductance of a silicon nitride nanopore**. Measured at concentrations of 1M (**a**), 100mM (**b**) and 10mM (**c**) KCl. The Conductance G is measured, whereas the temperature T and the surface charge σ are calculated according to the main text in order to estimate the necessary temperature or surface charge change to explain the observed data.*

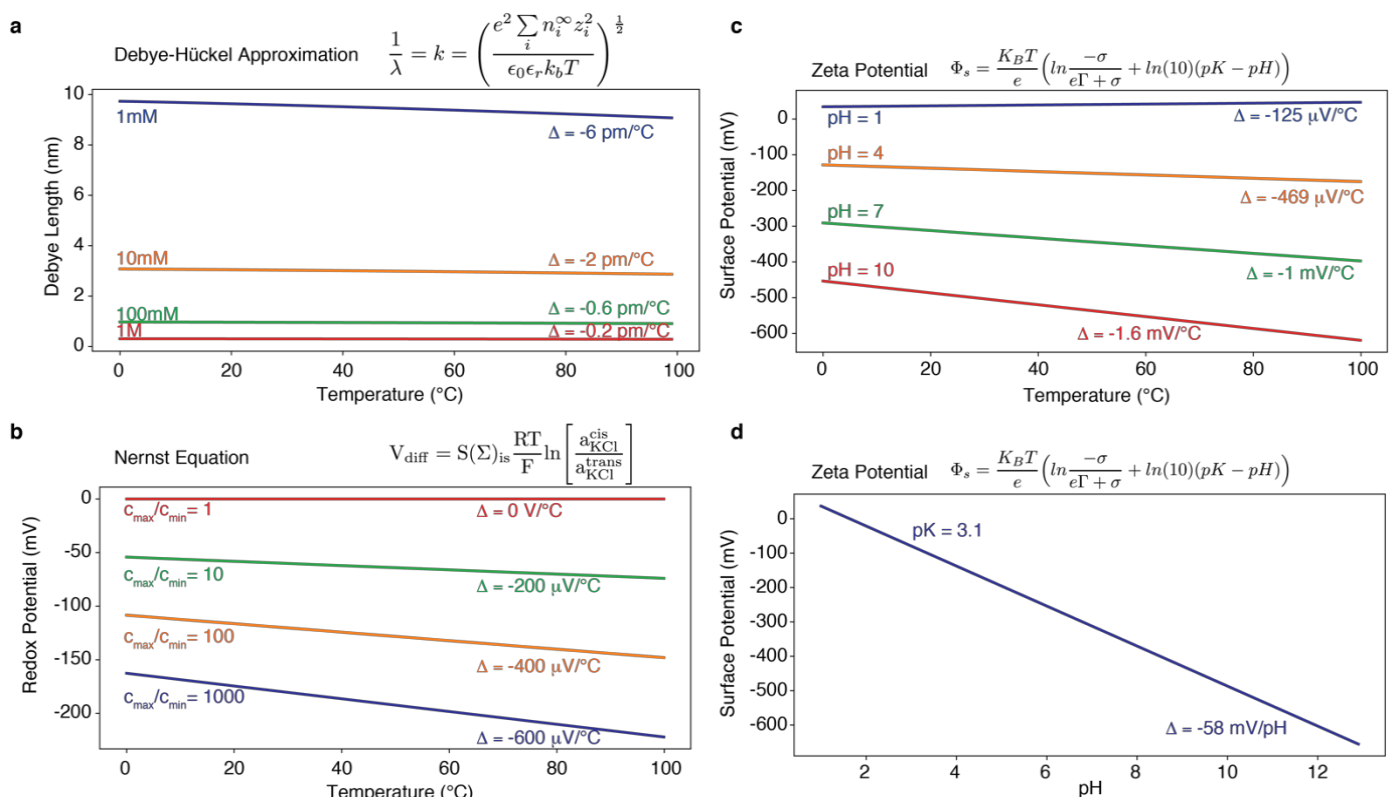

Figure S7: **Calculation of heat-dependent variables. a,** Debye Length vs temperature for different ionic dilutions. The data has been calculated using the Debye-Hückel approximation[7] and an analytical estimation of the dielectric constant of water[8]. **b,** Nernst potential vs. temperature for different concentration gradients. **c,** The surface potential vs temperature for different pH values. **d,** The surface potential as a function of pH for assuming a pK of 3.1. The surface potential as a function of pH for assuming a pK of 3.1.

# Tables

| Programmed Power [mW] | Real Power 643nm [mW] | Real Power 475nm [mW] |
|---|---|---|
| **150** | 135.9 | - |
| **140** | 127.1 | - |
| **130** | 117.9 | - |
| **120** | 109.1 | - |
| **100** | 91.7 | 98.3 |
| **50** | 47.1 | 49.1 |
| **20** | 19.89 | 19.6 |
| **10** | 10.7 | 9.76 |
| **1** | 1.8 | - |

Table S1: The laser power programmed through the software vs. the measured laser power

| Salt Concentration | Absolute Measured Potential | Potential Difference (with respect to 1M) | Theoretical |
|---|---|---|---|
| **1M** | 31.7mV | 0mV | 0mV |
| **10mM** | 82.9mV | 51.2mV | 58.6mV |
| **100mM** | 135mV | 103.3mV | 117.13mV |
| **1mM** | 185mV | 153.3mV | 175.7mV |

Table S2: Measurement of the redox potential at the Ag/AgCl electrodes

| 10nm pore | 643nm | | 475nm | |
|---|---|---|---|---|
| | Dark | 752 uW/cm$^2$ | Dark | 1.3mW/cm$^2$ |
| Osmotic Current | 1.41 nA | 1.63 nA | 1.32 nA | 1.48 nA |
| Osmotic Voltage | 95.64 mV | 98.63 mV | 92.30 mV | 95.37 mV |
| Osmotic Power | 134.72 pW | 160.85 pW | 122.09 pW | 140.66 pW |

| 3nm pore | 643nm | | 475nm | |
|---|---|---|---|---|
| | Dark | 752 uW/cm$^2$ | Dark | 1.3mW/cm$^2$ |
| Osmotic Current | 172.76 pA | 291.34 pA | 207.99 pA | 283.98 pA |
| Osmotic Voltage | 71.03 mV | 98.20 mV | 73.21 mV | 91.86 mV |
| Osmotic Power | 12.41 pW | 28.67 pW | 15.33 pW | 26.27 pW |

*Table S3: Summary of the measured osmotic parameters in Figure 2*